\documentclass[useAMS,usenatbib,usegraphicx]{mn2e}
\title[Formation history of PAHs in galaxies]{Formation History of Polycyclic Aromatic Hydrocarbons in Galaxies}
\author[J. Y. Seok, H. Hirashita, and R. S. Asano]{Ji Yeon Seok$^{1}$\thanks{E-mail:jyseok@asiaa.sinica.edu.tw}, Hiroyuki Hirashita$^1$, and Ryosuke S. Asano$^2$ \\
$^1$Institute for Astronomy and Astrophysics, Academia Sinica, P. O. Box 23-141, Taipei 10617, Taiwan\\
$^2$Department of Particle and Astrophysical Science, Nagoya University, Furo-cho, Chikusa-ku, Nagoya 464-8602, Japan}

\date{Submitted}

\pagerange{\pageref{firstpage}--\pageref{lastpage}} \pubyear{xxxx}

\begin{document}

\label{firstpage}
\maketitle

\begin{abstract}
Polycyclic aromatic hydrocarbons (PAHs) are some of the major dust components in the interstellar medium (ISM). We present our evolution models for the abundance of PAHs in the ISM on a galaxy-evolution timescale. We consider shattering of carbonaceous dust grains in interstellar turbulence as the formation mechanism of PAHs while
the PAH abundance can be reduced by coagulation onto dust grains, destruction by supernova shocks, and incorporation into stars. We implement these processes in a one-zone chemical evolution model to obtain the evolution of the PAH abundance in a galaxy.
We find that PAH formation becomes accelerated above certain metallicity where shattering becomes efficient.
For PAH destruction, while supernova shock is the primary mechanism in the metal-poor environment,
coagulation is dominant in the metal-rich environment. We compare the evolution of the PAH abundances in our models with observed abundances in galaxies with a wide metallicity range.
Our models reproduce both the paucity of PAH detection in low metallicity galaxies and the metallicity-dependence of the PAH abundance in high-metallicity galaxies. The strong metallicity dependence of PAH abundance appears as a result of the strong metallicity dependence of the dust mass increase by the accretion of metals onto dust grains, which are eventually shattered into PAHs. We conclude that the observational trend of the PAH abundance can be a natural consequence of shattering of carbonaceous grains being the source of PAHs.
To establish our scenario of PAH formation, observational evidence of PAH formation by shattering would be crucial.

\end{abstract}


\section{INTRODUCTION}

Since unidentified infrared emission (UIE) bands were discovered, the UIE bands have been seen in various astrophysical objects, which turned out to be abundant and ubiquitous in space. The UIE bands are commonly attributed to polycyclic aromatic hydrocarbons \citep[PAHs,][]{leger84,alla85,li12} among various UIE carrier candidates that have been proposed such as mixed aromatic/aliphatic organic nano particles \citep[MAONs,][]{kwok13} or hydrogenated amorphous (nano-) particles \citep[a-C:H,][]{jones09,jones12}. Adopting PAHs as a main carrier, PAHs are considered to be an important component in the interstellar medium (ISM) and play a key role in the energy balance via absorbing ultraviolet (UV) photons and emitting in infrared and the ionization balance via interaction with electrons and ions \citep[for a recent review, see][]{tiel08}. 

A number of infrared (IR) observations have been carried out toward various types of galaxies, which show rich PAH features in many cases. Measured intensities of PAH bands show strong metallicity-dependence \citep[e.g.,][]{engel05,madden06}. In particular, a dearth of PAH features is seen in low-metallicity galaxies with active star-formation such as blue compact dwarf galaxies \citep[BCDs; e.g.,][]{hunt10}. This is noticeable in the context that the strength of PAH bands are widely used as an indicator of star formation in galaxies \citep[e.g.,][]{forster04} and that mid-infrared (MIR) emission from starburst galaxies is often overwhelmed by PAH emission.

The paucity of PAH emission in low-metallicity galaxies can result from either efficient destruction 
and/or intrinsic deficiency of PAHs. 
UV radiation from massive stars may play an important role in PAH destruction 
in low metallicity galaxies probably because of the hardness of the radiation \citep{plante02}. 
Also, based on an observational indication of enhanced supernova (SN) activities in low-metallicity starbursts, \citet{ohall06} considered that PAH destruction by SN shocks is the dominant reason for the lack of PAHs. 
\citet{ohall06} showed that the PAH abundance is correlated more strongly with a SN shock indicator than with a hardness indicator, 
so they concluded that SN shocks are of larger significance to PAH destruction than UV radiation.

On the other hand, young galaxies that have not had sufficient time for asymptotic giant branch (AGB) stars to produce PAHs and to supply them into the ISM would have low PAH abundances \citep{gal08}. 
Although carbon-rich AGB stars are considered as a supplier of interstellar PAHs \citep[e.g.,][]{latter91}, however,
few carbon-rich AGB stars have shown direct observational evidence for the PAH emission \citep[e.g.,][]{buss91}. 
Moreover, \citet{cher92} calculated PAH formation yields in carbon-rich stellar outflows, which are much smaller than values inferred from the observed dust content of late-type carbon-rich stellar envelopes.
They concluded that it is unlikely that PAHs are produced in the stellar outflow itself. 
Furthermore, since the metallicity does not necessarily reflect the age \citep[e.g.,][]{kunth00},
it may not be straightforward to explain the clear metallicity trend of PAH abundance by the dominance of supernova destruction and/or the lack of AGB stars, both of which depend strongly on the age.

A different approach to PAH formation was suggested by \citet{jones96} via shattering in grain-grain collisions \citep*[see also][]{guillet11}. Large grains fragment into small grains by shattering in grain-grain collisions in interstellar shocks, and hence a lot of small grains are produced by this shattering process. \citet{jones96} calculated that about 5 to 15\% of the initial graphite grain mass in interstellar shocks with shock velocities $v_\mathrm{shock}\sim50-200$ km s$^{-1}$ might become sub-14 \AA~fragments which correspond the size range of PAHs. More recently, \citet{micel10} superseded the earlier work by a detailed study of PAH processing in interstellar shocks ($v_{\rm shock}\sim50-200$ km s$^{-1}$). Interstellar PAHs are completely destroyed by shocks with velocities greater than $\sim100$ km s$^{-1}$ whereas PAHs are not completely destroyed even though they lose a fraction of carbon atoms for shocks slower than 100 km s$^{-1}$. 
Later, the effect of grain shattering on the dust size distribution by C-type shocks ($v_{\rm shock}\sim20-40$ km s$^{-1}$) was studied by \citet{guillet11} who modeled gas-grain and grain-grain processes in dense clouds ($n_{\rm H}\ge 10^4$ cm$^{-3}$, where $n_\mathrm{H}$ is the number density of hydrogen nuclei). They showed that numerous small grains are produced in the shock through the fragmentation of large grains although it is only significant where densities are higher than $\sim10^5$ cm$^{-3}$.

The importance of shattering as a source of small grains is further pronounced by
\citet{hiro09}, who showed that shattering generally occurs in the diffuse ISM as long
as the medium is turbulent. The grains obtain moderate velocities up to $\sim10$ km s$^{-1}$
\citep{yan04}. Because of the velocities much smaller than $\sim100$ km s$^{-1}$,
the PAHs are expected to survive without being destroyed.
Therefore, it is worth considering shattering in turbulence as an
ubiquitous mechanism of PAH formation (or small grain formation in general).

In this paper, we aim at putting forward this hypothesis that PAHs form through
the fragmentations of large carbonaceous grains, by 
proposing a novel framework that describes
the evolution of small grain abundance. 
We particularly focus on PAHs as small grain species 
and clarify what kind of evolutionary features can be predicted by the models 
for the PAH abundance. Indeed, as shown later, our models
naturally explain the strong metallicity dependence of the PAH abundance for the
first time.

This paper is organized as follows. We describe our PAH evolution models in section 2 and present the calculation results for the evolution of the PAH abundance in galaxies in section 3. The results are discussed in section 4. Finally, section 5 gives the conclusions. \\

\section{PAH evolution model in galaxies}

We model the evolution of the total PAH mass in a galaxy. The main purpose of our modeling is to examine the hypothesis that PAHs (or PAH-like carbonaceous fragments) form through the collisional fragmentation of carbonaceous grains \citep[shattering,][]{jones96}. Recently, \citet{jones09}
proposed that hydrogenated amorphous carbon (a-C:H), which may be a
viable candidate for the interstellar carbonaceous dust, can be precursors of PAHs
as a result of progressive aromatization and photo-fragmentation.

There are two major sites for shattering in the ISM: 
One is in SN shocks, and the other is in turbulence in the diffuse ISM \citep{yan04,hiro09}. 
While the grain velocities in SN shocks are so high that formed PAHs are likely to be completely destroyed \citep{micel10}, 
those in turbulence are moderated \citep[$\sim10$ km s$^{-1}$,][]{yan04} so PAHs can survive. 
Therefore, in this work, we ``revisit" \citet{jones09}'s idea about PAH formation through fragments under a new motivation of turbulence-driven shattering \citep{hiro09}. For the formation mechanism of PAHs, we concentrate on shattering of carbonaceous grains and do not take other possible PAH formation mechanisms such as PAH formation in stellar winds into account. Comparison with other possible PAH formation mechanisms is discussed in Section \ref{sec:dis_other}. For the destruction or loss mechanisms, we consider coagulation of PAHs into larger grains, destruction by SN shocks, and incorporation into star formation (astration).


The time evolution of the total PAH mass in the galaxy is then
described as
\begin{equation}
\frac{dM_{\rm PAH}(t)}{dt}=-\mathcal{P}(t){\rm SFR}(t)-\frac{M_{\rm PAH}}{\tau \prime _{\rm SN}}+\frac{M_{\rm c,dust}}{\tau _{\rm shat}}-\frac{M_{\rm PAH}}{\tau _{\rm coag}},
\label{eq:dmdt}
\end{equation}
where $M_{\rm PAH}$ is the total mass of PAHs, $\mathcal{P}$ is the PAH-to-gas mass ratio ($\mathcal{P}=M_{\rm PAH}/M_{\rm ISM}$ where $M_{\rm ISM}$ is the total mass of the ISM), SFR is the star formation rate, and $M_{\rm c,dust}$ is the total mass of carbonaceous grains. Three timescales are introduced: $\tau\prime_{\rm SN}$\footnote{$\prime$ is used to distinguish the PAH destruction timescale from the dust destruction timescale in \citet{asano12}.}, $\tau_{\rm shat}$, and $\tau_{\rm coag}$ are the timescales of PAH destruction in shocks, shattering, and coagulation, respectively. The SFR is evaluated according to the Schmidt law \citep[SFR $\propto M_{\rm ISM}^n$,][]{schm59} with $n=1$, given as 
\begin{equation}
{\rm SFR}(t)=\frac{M_{\rm ISM}(t)}{\tau_{\rm SF}},
\end{equation}
where $\tau_{\rm SF}$ is the star formation timescale
(we give $\tau_\mathrm{SF}$ below). Determination of
$\tau\prime_\mathrm{SN}$, $\tau_\mathrm{shat}$, and $\tau_\mathrm{coag}$
is described in the subsections below. We calculate $M_\mathrm{c,dust}$ by using the
models in \citet{asano12}.

Here we briefly review the calculation of total dust content in \citet{asano12}. They formulate the evolution of the total dust mass ($M_\mathrm{dust}$) by using a chemical
evolution framework; that is, they consider the time evolution of the
total masses of gas, metals, and dust in a galaxy, in a way consistent with
the stellar evolution and the ISM recycling. For the evolution of dust mass,
they consider dust supply from AGB stars and SNe, and dust mass growth by the accretion of
metals onto dust grains in the ISM (simply called dust growth in this paper),
as the
sources of dust, and SN shocks as the destruction mechanism of dust.
The time evolution of carbonaceous dust mass
$M_\mathrm{c,dust}$ is derived by replacing the total dust mass
with the carbonaceous dust mass in equation (4) of \citet{asano12} and
considering only the contribution from carbonaceous dust for the stellar dust production.
We adopt $\eta =0.5$ (the mass fraction of the clouds
hosting dust growth; i.e., the cold neutral medium, molecular clouds, and dense clouds; see Table \ref{tab:ism}). As shown by \citet{asano12},
the metallicity at which dust growth
becomes prominent scales with $(\eta\tau_\mathrm{SF})^{-1/2}$. Since the
change of $\eta$ is degenerate with that of $\tau_\mathrm{SF}$, we concentrate
on the variation of $\tau_\mathrm{SF}$. We examine
$\tau_\mathrm{SF}=0.5$, 5, and 50 Gyr to cover typical
star formation timescales seen in nearby star-forming galaxies
\citep[e.g.,][]{kenn98}.
 We assume the total baryonic mass
(i.e., the initial ISM mass) to be $10^{10}~M_\odot$; all the total dust, PAH, metal, and gas
masses just scale with the total baryonic mass, so that the dust-to-gas ratio,
the PAH-to-gas ratio, and metallicity do not depend on the
total baryonic mass adopted.


\subsection{Shattering and coagulation}\label{sec:shco}

To derive $\tau_{\rm shat}$ and $\tau_{\rm coag}$, 
we first investigate how a grain size distribution of carbonaceous dust, 
$n_\mathrm{c}(a)$, alters due to shattering and coagulation. 
As for the initial size distribution, we adopt two cases.
One is a simple power-law distribution with enhancement at the size range of PAHs. 
This is basically based on the size distribution by \citet*[hereafter MRN]{mrn77}, 
which is derived from the observed interstellar extinction curve. 
As the range of grain radius in MRN is roughly 
between 0.005 \micron~(50 \AA) and 0.25 \micron~for graphite, 
we extrapolate the MRN distribution down to the size range of PAHs 
with a normalization consistent with the observed PAH abundance (see details below). 
This size distribution is referred to as ``MRN+enhanced PAH''.
The other is the size distribution by \citet[hereafter WD01]{wein01} 
who attempted to fit the extinction curves more precisely. 
WD01 include the sum of two log-normal size distributions 
taking the very small grain population including PAHs into account. 
The fraction of carbon in the log-normal components is assumed to be
$b_{\rm C}=6\times10^{-5}$, which is favored by \citet{li01} from their
analysis of infrared emission in the Milky Way. For the other parameters,
see Table 1 in WD01.

In the following, we summarize how to build the initial size distribution with a power-law (i.e., MRN+enhanced PAH). 
Following MRN, $n_\mathrm{c}(a)\, da$, 
the number density of carbonaceous grains (excluding PAHs) with radii between $a$ and $a+da$, is given by
\begin{equation}
n_{\rm c}(a)=\mathcal{C}_{\rm c}a^{-3.5}~(a_{\rm min,c}\le a\le a_{\rm max,c}),
\label{eq:n_c}
\end{equation}
where $\mathcal{C}_{\rm c}$ is the normalizing constant, and $a_\mathrm{min,c}$ and $a_\mathrm{max,c}$ are the minimum and maximum radii for the carbonaceous dust grains, respectively.
We adopt $a_{\rm max,c}=0.25$ \micron~from MRN and define $a_{\rm min,c}=a_{\rm max,PAH}$, where $a_\mathrm{max,PAH}$ is the maximum radius of PAHs (see next).
For the PAH size distribution, $n_\mathrm{PAH}(a)$, we apply the same power-law with a
different normalization:
\begin{equation}
n_{\rm PAH}(a)=\mathcal{C}_{\rm PAH}a^{-3.5}~(a_{\rm min,PAH}\le a\le a_{\rm max,PAH}),
\label{eq:n_pah}
\end{equation}
where $\mathcal{C}_{\rm PAH}$ is the normalizing constant. 
Since the size of typical PAHs with $N_{\rm C}=20-100$
($N_{\rm C}$: the number of carbon atoms) is about a few \AA, and PAH clusters ($N_{\rm C}=100-1000$) can be up to $\sim20$ \AA~\citep{tiel08}, we set ($a_{\rm min,PAH}, a_{\rm max, PAH}$) = (0.0003 \micron, 0.002 \micron). 
The power index adopted for the PAH size distribution has only a minor influence on the timescales.
As shown later (Figure \ref{fig:dtog_wd01}), the initial size distributions 
do not affect the following results significantly,
and indeed, the total abundance of PAHs is further more important.

\begin{figure}
\includegraphics[width=86mm]{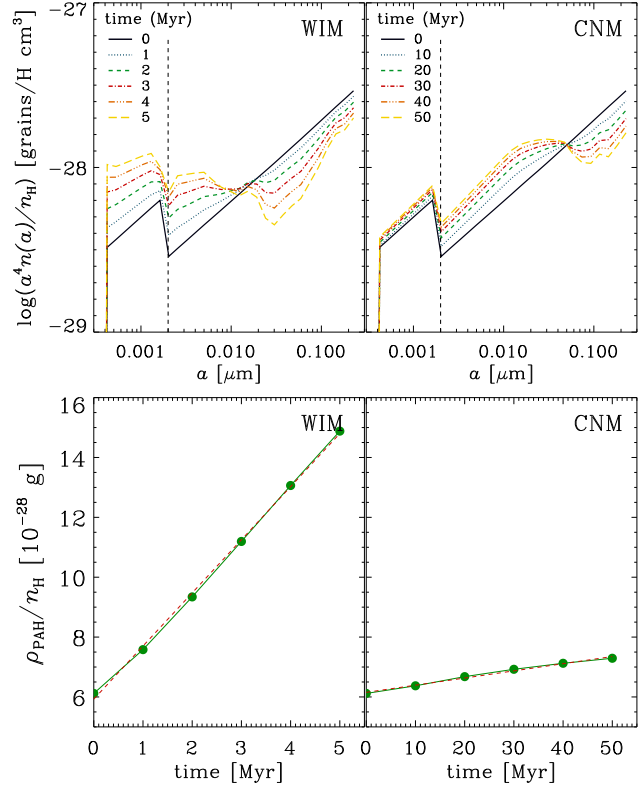}
\caption{Top: Grain size distributions of carbonaceous dust including PAHs in WIM and CNM, 
where shattering is effective, for the MRN+enhanced PAH size distribution. 
The initial size distribution is shown by a solid line (see section \ref{sec:shco}). 
The different lines present the size distributions for different durations of shattering
as indicated in each panel. 
The grain size distribution is expressed by multiplying $a^4$ to show the mass distribution per logarithmic grain radius and is normalized to the hydrogen density ($n_{\rm H}$). 
The maximum radius of PAHs is denoted by the dashed line (20 \AA). 
Bottom: The evolution of the total PAH mass density per hydrogen by shattering in WIM and CNM.
A linear fit, which is used to derive the shattering timescale, is overlaid (dashed line). }
\label{fig:shat}
\end{figure}

\begin{figure}
\includegraphics[width=86mm]{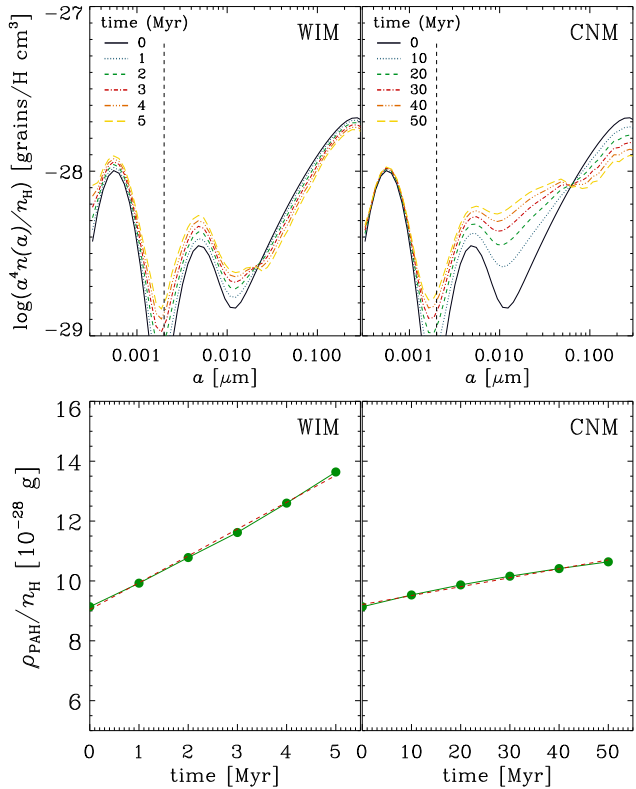}
\caption{Same as Figure \ref{fig:shat} calculated for the size distribution of \citet{wein01}. 
\label{fig:shat_wd01}}
\end{figure}

Assuming that grains are spherical with a constant material density (we adopt the material density of graphite, $\rho_{\rm gr}=2.2$ g cm$^{-3}$ for carbonaceous dust; \citealt{draine84}),
the normalization factor $\mathcal{C}=\mathcal{C}_\mathrm{c}$ or $\mathcal{C}_\mathrm{PAH}$ is determined by
\begin{equation}
\mathcal{R}m_{\rm H}n_{\rm H}=\int^{a_{\rm max}}_{a_{\rm min}}\frac{4\pi}{3}a^3\rho_{\rm gr}\mathcal{C}a^{-3.5} da,
\end{equation}
where $(a_\mathrm{min},\, a_\mathrm{max})=(a_\mathrm{min,c},\, a_\mathrm{max,c})$ and ($a_\mathrm{min,PAH}$, $a_\mathrm{max,PAH}$) for carbonaceous dust and PAHs, respectively, $n_{\rm H}$ is the hydrogen number density in each medium (Table \ref{tab:ism}), $m_{\rm H}$ is the hydrogen atom mass, $\mathcal{R}$ is the dust (PAH) abundance relative to hydrogen. 
We refer to $\mathcal{R}=3.4\times10^{-3}$ ($4\times10^{-4}$) 
for carbonaceous dust (PAH) for the Milky Way \citep{takagi03}, 
which we adopt as representative values in solar-metallicity environments. 
With $\mathcal{C}_\mathrm{c}$ or $\mathcal{C}_\mathrm{PAH}$ derived above, 
the initial size distribution of MRN+enhanced PAH is shown 
in Figures \ref{fig:shat} and \ref{fig:coag} with solid lines.
The PAH abundance is enhanced compared to the extrapolation of $n_\mathrm{c}(a)$.
This size distribution is similar to the size distributions 
for more elaborate dust models by \citet{zub04}. In addition, 
the size distribution of WD01 is shown at Figures \ref{fig:shat_wd01} and \ref{fig:coag_wd01}.

In the following calculations of shattering and coagulation, we apply
the bulk properties of graphite to both the PAH and carbonaceous dust
components, since graphite is the most commonly applied species in
considering the carbonaceous components of solid particles \citep[e.g.,][]{draine84}. 
In fact, the details in the treatment of PAHs are not essential for our purpose 
of investigating a possibility of PAH formation by shattering of carbonaceous dust 
(see Section \ref{sec:unc} for further discussion on the adopted bulk properties).

Under the two initial size distributions, we calculate the
change of the grain size distributions by shattering and coagulation, adopting
the formulation in \citet{hiro09}. We outline their models in the following. 
\citet{hiro09} solved shattering and coagulation
equations under size-dependent grain velocity dispersions calculated by an analytic
method developed by \citet{yan04}, who considered that the grains acquire
velocity dispersions by gas drag and
gyroresonance in turbulent medium. As shown in Table \ref{tab:ism}, we consider
all the ISM phases investigated by \citet{yan04}: the cold neutral medium
(CNM), warm neutral medium (WNM), warm ionized medium (WIM),
molecular clouds (MC), and dense clouds (DC). Following
\citet{yan04}, we consider two cases (DC1 and DC2) for DC to examine the
uncertainty in the ionization degree.

\begin{table*}
\begin{minipage}{140mm}
\begin{center}
\caption{Physical properties of ISM phases \label{tab:ism}}
	\begin{tabular}{@{}ccccccc@{}}
	\hline
	ISM phase & CNM & WNM & WIM & MC & DC1 & DC2 \\
	\hline
	$T$ (K) & 100 & 6000 & 8000 & 25 & \multicolumn{2}{c}{10} \\
	$n_{\rm H}$ (cm$^{-3}$) & 30 & 0.3 & 0.1 & 300 & \multicolumn{2}{c}{$10^4$} \\
	$n_{\rm e}$ (cm$^{-3}$) & 0.03 & 0.03 & 0.0991 & 0.03 & 0.01 & 0.001 \\
	Mass fraction (\%) & 24 & 36 & 23 & 6 & \multicolumn{2}{c}{11} \\
	& 30 & 38 & 14 & \multicolumn{3}{c}{18} \\
	\hline
	\end{tabular}
\end{center}
   \medskip
   
$Note$: $T$ is the gas temperature, $n_{\rm H}$ is the hydrogen number density, and $n_{\rm e}$ is the number density of electrons. The ISM phases listed are the cold neutral medium (CNM), warm neutral medium (WNM), warm ionized medium (WIM), molecular clouds (MC), and dense clouds (DC). For DC, we consider two cases for the ionization degree. All parameters but the mass fraction of ISM are from \citet{yan04}. The mass fractions in the forth and fifth rows are derived from the mass of \textsc{H ii}, \textsc{H i}, and H$_2$ in the Milky Way with the density-weighted volume filling factor, $n_{\rm H}f_V$ in \citet[]{draine11} and from the total mass of each ISM phase in \citet{tiel05}, respectively. The total mass of MC and DC are not specified separately in \citet{tiel05}, so we divide the total mass of the clouds into MC and DC with the same ratio in \citet{draine11} for further analysis  (i.e., MC $\simeq6\%$ and DC $\simeq12\%$). CNM: cold neutral medium, WNM: warm neutral medium, WIM: warm ionized medium, MC: molecular cloud, and DC: dark cloud. 
\end{minipage} 
\end{table*}

As shattering or coagulation takes place in various ISM phases, the grain size distributions evolve differently. 
The resulting grain size distributions are shown in the top panels of Figures \ref{fig:shat}--\ref{fig:coag_wd01}, 
where the size distribution is expressed by multiplying $a^4$ to show the mass distribution in each logarithmic bin of the grain radius. 
We discretize the logarithmic grain radius into 32 bins and 
confirm that the number of bins does not affect the results \citep{hiro09}. 
Then, the volume mass densities of PAHs and carbonaceous dust,
$\rho_{\rm PAH}$ and $\rho_{\rm c,dust}$, are given by
\begin{equation}
\rho_{\rm PAH}=\int^{a_{\rm max,PAH}}_{a_{\rm min,PAH}}\frac{4\pi}{3}a^3\rho_{\rm gr}n_{\rm PAH}(a) da
\end{equation}
and
\begin{equation}
\rho_{\rm c, dust}=\int^{a_{\rm max,c}}_{a_{\rm min,c}}\frac{4\pi}{3}a^3\rho_{\rm gr}n_{\rm c}(a) da.
\end{equation}
The evolution of $\rho_{\rm PAH}$ is shown in the bottom panels of Figures 
\ref{fig:shat}--\ref{fig:coag_wd01}. In the case of the WD01 size distribution, 
we assume that the size distribution of particles with $a\leq0.002~\micron$ 
correspond to $n_{\rm PAH}$ and derive $\rho_{\rm PAH}$ and $\rho_{\rm c, dust}$ 
(Figures \ref{fig:shat_wd01} and \ref{fig:coag_wd01}).

Figures \ref{fig:shat} and \ref{fig:shat_wd01} show that shattering changes the size distribution in WIM most efficiently. This is because grains larger than a few $\times10^{-6}$ cm are easily accelerated above the shattering threshold velocity ($v_{\rm shat}=1.2$ km s$^{-1}$ for graphite) by gyroresonance. Shattering also alters the grain size distribution in CNM, but the variation of the total PAH mass is insignificant.\footnote{\citet{hiro09} mentioned that the shattering in CNM is not conclusive because a slight modification of the shattering threshold velocity causes a significant variation of the resulting grain size distribution.}
In the other ISM phases, shattering does not occur because grain velocities are
smaller than the shattering threshold.
On the other hand, coagulation occurs mainly in MC and DC, and the largest variation is seen in DC because of high density (Figures \ref{fig:coag} and \ref{fig:coag_wd01}). 
Due to the different efficiency of shattering/coagulation in each ISM phase, 
$\rho_{\rm PAH}$ and $\rho_{\rm c,dust}$ evolves on various timescales 
(the bottom panels of Figures \ref{fig:shat}--\ref{fig:coag_wd01}). 
Based on the change of $\rho_\mathrm{PAH}$, we estimate $\tau_\mathrm{shat}$ 
and $\tau_\mathrm{coag}$ at the solar metallicity and later scale them with
the dust-to-gas ratio as explained in Sections \ref{sec:stime} and \ref{sec:ctime}.

\subsubsection{Shattering timescale}\label{sec:stime}

By using the time evolution of $\rho_\mathrm{PAH}$ by shattering,
$[d\rho_\mathrm{PAH}/dt]_\mathrm{shat}$, we evaluate the shattering
timescale in each ISM phase by
\begin{equation}
\bigg[\frac{d\rho_{\rm PAH}}{dt}\bigg]_{\rm shat}=\frac{\rho_{\rm c,dust}}{\tau_{{\rm shat,}i}^0},
\label{eq:drho_sh}
\end{equation}
where $\tau_{{\rm shat},i}^0$ is the shattering timescale in each ISM phase at the Milky Way dust-to-gas ratio
($i$ indicates the ISM phase). We fit the time variations of $\rho_{\rm PAH}$
with a linear slope to derive $\tau_{\mathrm{shat},i}^0$ (Figure \ref{fig:shat}, \textit{bottom}).
We assume $\rho_\mathrm{c,dust}$ to be constant, since the change is negligible.
The obtained values of $\tau_{{\rm shat},i}^0$ are shown only for CNM and WIM
in Table \ref{tab:tau0} as shattering is negligible in the other ISM phases.
The shattering timescale in each medium ($\tau_{{\rm shat,}i}$) is inversely proportional to the carbonaceous dust-to-gas ratio ($\mathcal{D}\equiv M_{\rm c,dust}/M_{\rm ISM}$) given as 
\begin{equation}
\tau_{{\rm shat},i}(\mathcal{D})\simeq\tau_{{\rm shat},i}^0\bigg(\frac{\mathcal{D}}{\mathcal{D}_\odot}\bigg)^{-1},
\label{eq2}
\end{equation}
where $\mathcal{D}$ is normalized to \citet{asano12}'s value at the solar metallicity
($\mathcal{D}_\odot=0.017$).
Their calculations return a systematically large carbonaceous dust-to-gas ratio because of the
data adopted for the dust and metal formation in AGB stars and SNe. In other words,
the current dust formation models in stellar sources have still a large uncertainty,
which should be improved by future development of dust and metal production calculations
in stars. However, as seen below, the increase of PAH abundance is mostly governed by the
increase of carbonaceous dust abundance by grain growth, and most models
commonly show the dust growth at a similar metallicity level
\citep[e.g.,][]{dwek98,zhukovska08,inoue11,kuo12}. Because of this
robust feature commonly seen in dust evolution models, 
our discussions below hold even if we adopt models other than \citet{asano12}.

We finally obtain the shattering timescale of the entire ISM ($\tau_{\rm shat}$) by weighted summation of $\tau_{{\rm shat},i}$ for the mass fraction of each ISM phase ($f_{{\rm mass},i}$): 
\begin{equation}
\tau_{\rm shat}(\mathcal{D})=\frac{1}{\sum_{i} f_{{\rm mass},i}/\tau_{{\rm shat,}i}}.
\label{eq:sh}
\end{equation}
We adopt $f_{{\rm mass},i}$ from \citet{draine11} and \citet{tiel05} listed in Table \ref{tab:ism}. 
Considering CNM and WIM, where shattering takes place, we obtain, 
for the MRN+enhanced PAH size distribution, $\tau_{\rm shat}(\mathcal{D}_\odot)\simeq1.3\times10^8$ yr and $2.1\times10^8$ yr at the solar metallicity in \citet{asano12} using the mass fractions in \citet{draine11} and those in \citet{tiel05}, respectively. 
For the WD01 size distribution, we obtain $\tau_{\rm shat}(\mathcal{D}_\odot)\simeq1.8\times10^8$ yr and $2.8\times10^8$ yr for the mass fractions in \citet{draine11} and those in \citet{tiel05}, respectively. 
Although $f_{{\rm mass},i}$ still remains uncertain, 
we find that the different mass fractions do not
change our results significantly. 
Adopting the mass fractions in \citet{draine11} as our standard case, 
we further use $\tau_\mathrm{shat}(\mathcal{D})=1.3 (1.8)\times 
10^8(\mathcal{D}/\mathcal{D}_\odot)^{-1}$ yr, for the MRN+enhanced 
PAH size distribution (WD01 size distribution) in this paper. 


\subsubsection{Coagulation timescale}\label{sec:ctime}

\begin{figure*}
\begin{minipage}{177mm}
\centering
\includegraphics[width=170mm]{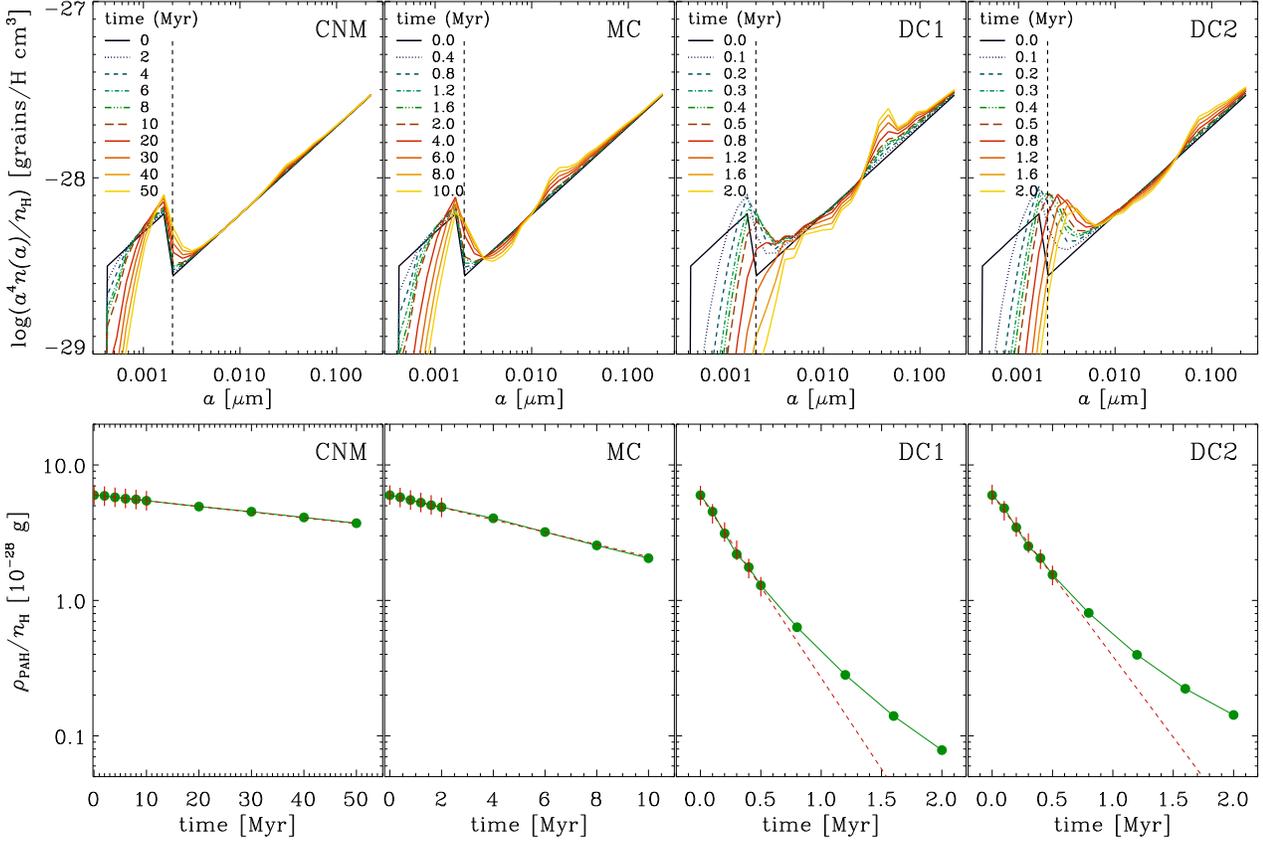}
\caption{Same as Figure \ref{fig:shat}, but the evolution is driven by coagulation. Top: The grain size distribution in CNM, MC, DC1, and DC2 where coagulation is effective. Bottom: The evolution of the total PAH mass by coagulation. We only use values at small $t$ for analytical fitting of an exponential function marked with \textit{bars} to derive the
coagulation timescale. 
\label{fig:coag}} 
\end{minipage}
\end{figure*}

\begin{figure*}
\begin{minipage}{177mm}
\centering
\includegraphics[width=170mm]{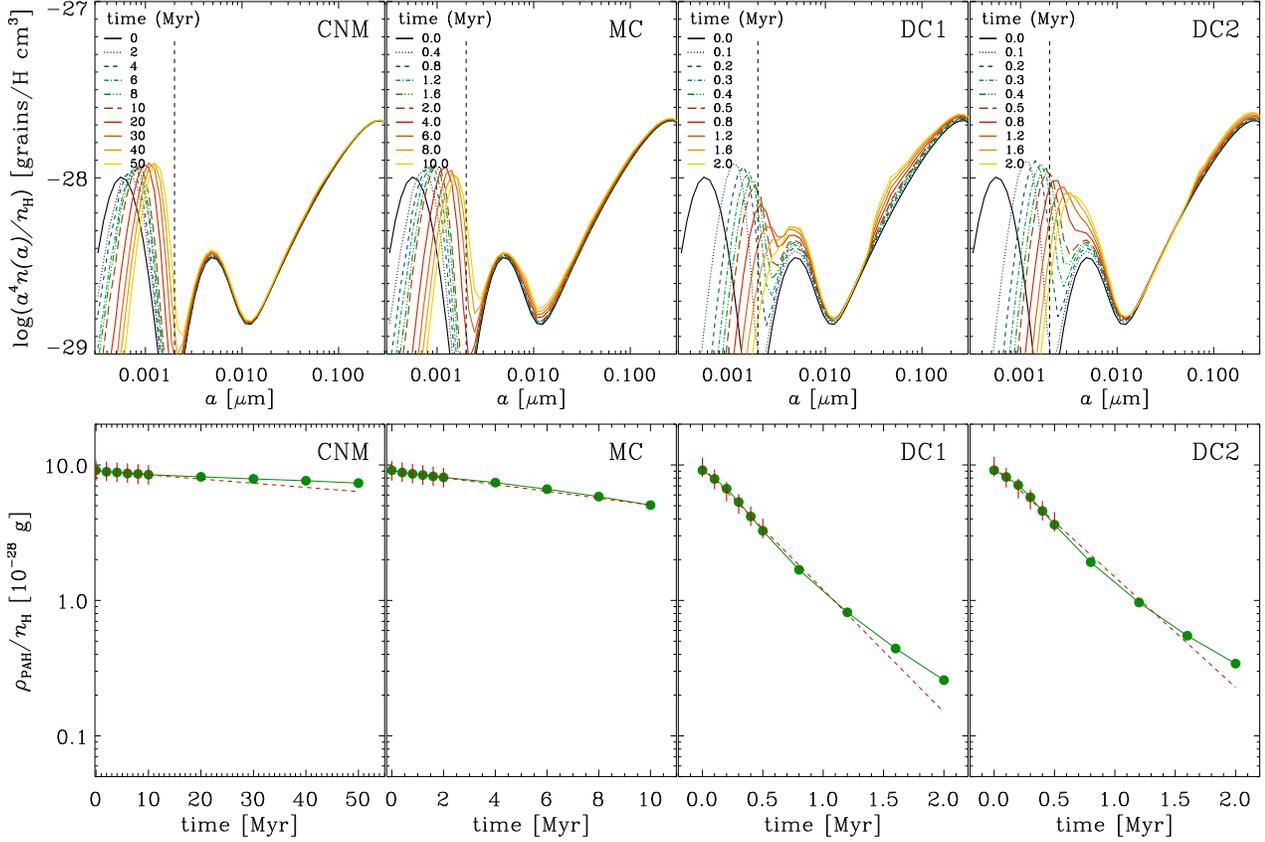}
\caption{Same as Figure \ref{fig:coag} calculated for the size distribution of \citet{wein01}. \label{fig:coag_wd01}} 
\end{minipage}
\end{figure*}

In a similar way, the timescale of coagulation in each ISM phase,
$\tau_{\mathrm{coag},i}^0$  can be evaluated by
\begin{equation}
-\bigg[\frac{d\rho_{\rm PAH}}{dt}\bigg]_{\rm coag}=\frac{\rho_{\rm PAH}}{\tau_{{\rm coag},i}^0},
\label{eq:drho_co}
\end{equation}
where $[d\rho_{\rm PAH}/dt]_{\rm coag}$ its the time evolution of $\rho_{\rm PAH}$ by coagulation.
Since this implies an exponential decay of $\rho_\mathrm{PAH}$, we fit the time variations of
$\rho_{\rm PAH}$ by coagulation with
$\rho_{\rm PAH}\propto e^{-t/\tau_{\mathrm{coag},i}^0}$ to determine
$\tau_{\mathrm{coag},i}^0$ (Figure \ref{fig:coag}, \textit{bottom}). As coagulation is not efficient in WIM and WNM, the fitting results of the other ISM phases are listed in Table \ref{tab:tau0}. As the timescales of coagulation in DC1 and DC2 are not much different from each other, we adopt DC1 for further estimation. The coagulation timescale is also inversely proportional to the dust-to-gas ratio ($\mathcal{D}$) given by
\begin{equation}
\tau_{{\rm coag,}i}(\mathcal{D})\simeq\tau_{{\rm coag,}i}^0\bigg(\frac{\mathcal{D}}{\mathcal{D}_\odot}\bigg)^{-1}.
\label{eq3}
\end{equation}
Note that the coagulated grains cannot be released into the interstellar space before the lifetime of the cloud although the coagulation inside the cloud is completed. Thus, when $\tau_{{\rm coag,}i}$
($i=$ DC)
is shorter than the lifetime of the DC ($\tau_{\rm DC}$), we substitute $\tau_{\rm coag, DC}$ with $\tau_{\rm DC}$. A typical lifetime of a DC is $\tau_{\rm DC}=3$--5 Myr \citep[and references therein]{berg07}, and we assume $\tau_{\rm DC}= 3$ Myr as our standard case and also examine
$\tau_\mathrm{DC}=1$ and 10 Myr. 
We derive the coagulation timescale of the entire ISM ($\tau_{\rm coag}$) in a similar
form to equation (\ref{eq:sh}):
\begin{equation}
\tau_{\rm coag}(\mathcal{D})=\frac{1}{\sum_{i} f_{{\rm mass},i}/\tau_{{\rm coag,}i}}.
\end{equation}
Taking CNM, MC, and DC1 into account, we obtain, for the MRN+enhanced PAH size distribution, 
\begin{eqnarray} \label{eq:tauc}
\tau_{\rm coag}(\mathcal{D}) \hspace{65mm} \nonumber\\
 = \left\{ \begin{array}{l}
2.9\times 10^6(\mathcal{D}/\mathcal{D}_\odot)^{-1} ~{\rm yr,}
 \hspace{5.6mm}\textrm{if $\tau_{\rm coag, DC1}$ $\geq\tau_{\rm DC}$,} \\ \\
10^7\times\{0.086(\mathcal{D}/\mathcal{D}_\odot)+1.1/(\tau_{\rm DC}/{\rm Myr})\}^{-1} ~{\rm yr,} \\
 \hspace{38.3mm}\textrm{if $\tau_{\rm coag, DC1}$ $<\tau_{\rm DC}$,} 
  \end{array} \right.
\end{eqnarray}
and, for the WD01 size distribution,
\begin{eqnarray} \label{eq:tauc_wd01}
\tau_{\rm coag}(\mathcal{D}) \hspace{65mm} \nonumber\\
 = \left\{ \begin{array}{l}
4.3\times 10^6(\mathcal{D}/\mathcal{D}_\odot)^{-1} ~{\rm yr,}
 \hspace{5.6mm}\textrm{if $\tau_{\rm coag, DC1}$ $\geq\tau_{\rm DC}$,} \\ \\
10^7\times\{0.052(\mathcal{D}/\mathcal{D}_\odot)+1.1/(\tau_{\rm DC}/{\rm Myr})\}^{-1} ~{\rm yr,} \\
 \hspace{38.3mm}\textrm{if $\tau_{\rm coag, DC1}$ $<\tau_{\rm DC}$.} 
  \end{array} \right.
\end{eqnarray}
Again, we adopt $f_{{\rm mass},i}$ in \citet{draine11}, and even if we follow \citet{tiel05}, it only changes $\tau_{\rm coag}(\mathcal{D})$ by less than 10\%.

\begin{table*}
\begin{minipage}{120mm}
\centering
   \caption{Shattering and coagulation timescales in each ISM phase at the solar metallicity \label{tab:tau0}}
	\begin{tabular}{@{}ccccccc@{}}
	\hline
	ISM phase & CNM & WNM & WIM & MC & DC1 & DC2 \\
	\hline	
	&\multicolumn{6}{c}{MRN+Enhanced PAH} \\
	\cline{2-7}
	$\tau_{{\rm shat},i}^0$ (Myr) & $2250\pm110$ & ---\footnote{Effect from shattering/coagulation is negligible.\label{note1}}	 & $30.3\pm0.6$  & ---$^{a}$ & ---$^{a}$ & ---$^{a}$ \\
	$\tau_{{\rm coag},i}^0$ (Myr) & $103\pm5$ & ---$^{a}$ & ---$^{a}$ & $9.54\pm0.23$ & $0.32\pm0.01$ & $0.36\pm0.01$ \\
	\hline	
	&\multicolumn{6}{c}{WD01} \\
	\cline{2-7}
	$\tau_{{\rm shat},i}^0$ (Myr) & $1307\pm67$ & ---$^{a}$ & $43.5\pm1.1$  & ---$^{a}$ & ---$^{a}$ & ---$^{a}$ \\
	$\tau_{{\rm coag},i}^0$ (Myr) & $140\pm12$ & ---$^{a}$ & ---$^{a}$ & $17.2\pm1.2$ & $0.48\pm0.02$ & $0.53\pm0.04$ \\
	\hline
	\end{tabular}

\end{minipage}   
\end{table*}


\subsection{Shock destruction}\label{sec:sn}

The timescale of PAH destruction by supernova shocks, $\tau\prime_{\rm SN}$, can be treated in the same way as dust destruction in shocks \citep{mckee89}. It is expressed as
\begin{equation}
\tau\prime_{\rm SN}=\frac{M_{\rm ISM}(t)}{\gamma_{\rm SN}(t)\int\varepsilon(v_s) dM_s(v_s)},
\end{equation}
where $\gamma_{\rm SN}$ is the SN rate, $\varepsilon(v_s)$ is the PAH destruction efficiency as a function of the shock velocity ($v_s$), and $M_s$ is the mass of the ISM swept up by a shock
with a shock velocity of $v_s$. \citet{jones94} derived $M_s(v_s)$ assuming a supernova remnant in the Sedov-Taylor stage with a ratio of warm to hot intercloud medium filling factor of $f_w/f_h =0.3/0.7 =0.43$, which yields $M_s(v_{s})=2914~M_\odot v^2_{s7}$ where $v_{s7}$ is shock velocity in units of $10^7$ cm s$^{-1}$ (i.e., 100 km s$^{-1}$). Then, the timescale $\tau\prime_{\rm SN}$ becomes
\begin{equation}\label{eq:tshock}
\tau\prime_{\rm SN}=\frac{M_{\rm ISM}(t)}{5828~\gamma_{\rm SN}(t)\int\varepsilon(v_{s7})/v_{s7}^3 dv_{s7}} \rm ~yr.
\end{equation}
Typical interstellar PAHs ($N_{\rm C}\sim20-100$) can be destroyed via either electron collisions or ion collisions, and small PAHs ($N_{\rm C}\simeq50$) are preferentially destroyed by electrons rather than ions \citep{micel10}. Following the analytical expressions for the destruction efficiency, $\varepsilon(v_{s7})$, at $0.5\le v_{s7}\le 2.0$ \citep[see Table 3 in][]{micel10}, we find $\int\varepsilon(v_{s7})/v_{s7}^3{\rm d} v_{s7} = 0.619$ for electron collisions.
With this value, we obtain the final expression for $\tau\prime_{\rm SN}$ from Equation (\ref{eq:tshock}),
\begin{equation}
\tau\prime_{\rm SN}=\frac{M_{\rm ISM}(t)}{3608~\gamma_{\rm SN}(t)} \rm ~yr,
\end{equation}
where $M_\mathrm{ISM}$ and $\gamma_\mathrm{SN}$ are given by
\citet{asano12}'s framework.

\section{Result}

\subsection{Evolution of PAH mass in galaxies}

\begin{figure*}
\begin{minipage}{177mm}
\centering
\includegraphics[width=170mm]{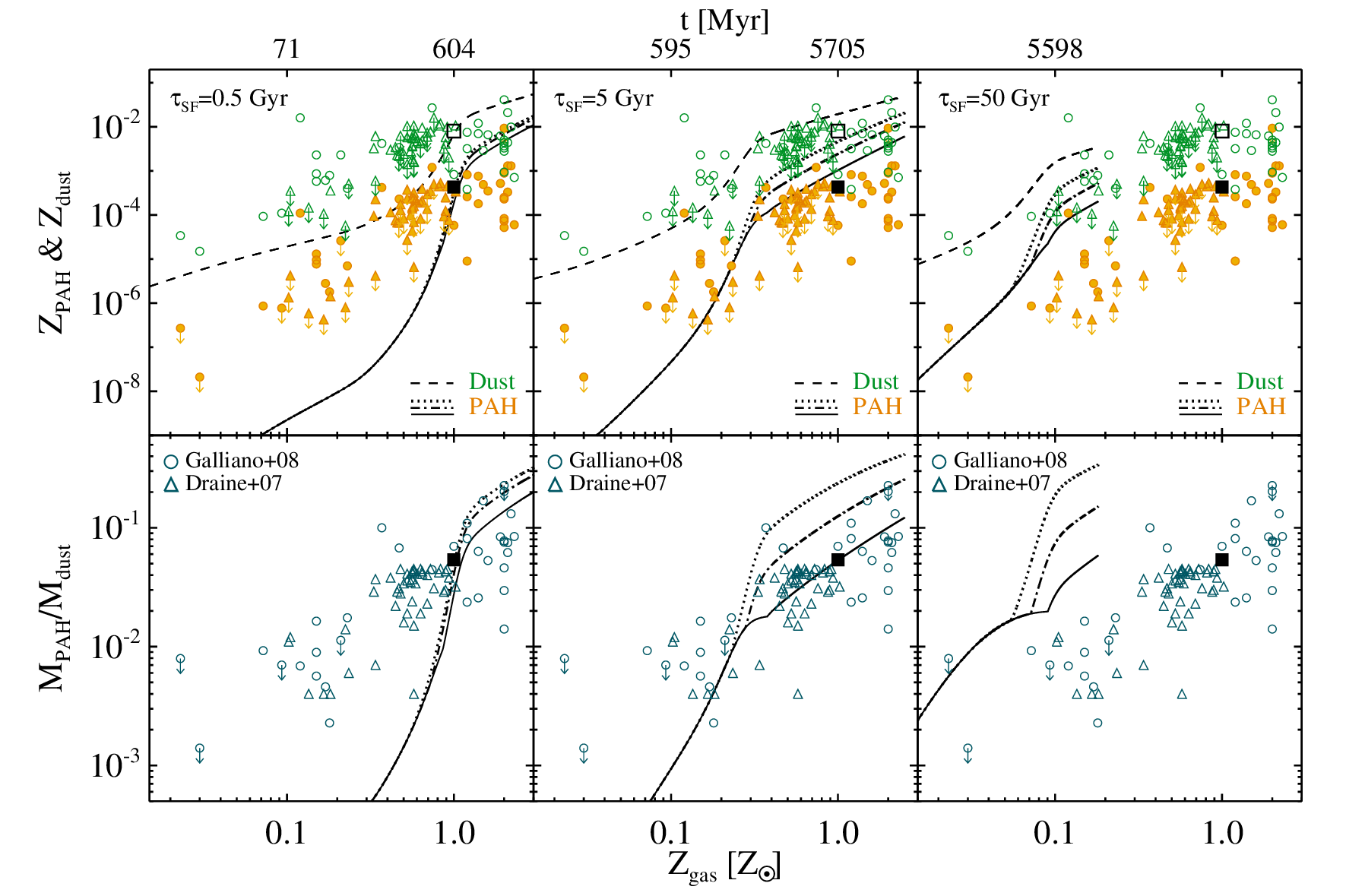}
\caption{Top: Calculated PAH-to-gas mass ratio ($Z_{\rm PAH}$) and dust-to-gas ratio 
($Z_{\rm dust}$, dashed line) as a function of metallicity ($Z_{\rm gas}$) for
$\tau_{\rm SF}=0.5$, 5, and 50 Gyr. 
As for $Z_{\rm PAH}$, the cases for $\tau_{\rm DC}=1$, 3, and 10 Myr are overlaid in each panel with solid, dash-dotted, and dotted lines, respectively. 
Observational data from \citet{gal08} and \citet{draine07} are overlaid for comparison 
(circles and triangles, respectively). $Z_{\rm dust}$ and $Z_{\rm PAH}$ are distinguished 
with open and filled symbols. Also, $Z_{\rm PAH}$ and $Z_{\rm dust}$ of the diffuse 
Galactic ISM are denoted with filled and open squares, respectively \citep{zub04}. 
The time corresponding to the metallicity are indicated on the 
top axis. As the model calculation terminates at 10 Gyr, the time at $Z_{\rm gas}=1$ cannot be 
marked in the case of $\tau_{\rm SF}=50$ Gyr. 
Bottom: Total mass ratio of PAH to dust ($Z_{\rm PAH}/Z_{\rm dust}$) for
$\tau_{\rm SF}=0.5$, 5, and 50 Gyr. 
\label{fig:dtog_Z}} 
\end{minipage}
\end{figure*}

In Figure \ref{fig:dtog_Z} ({\it top}), we show the variation of the PAH and dust abundances ($Z_{\rm PAH}\equiv M_\mathrm{PAH}/M_\mathrm{ISM}$ and $Z_{\rm dust}\equiv M_\mathrm{dust}/M_\mathrm{ISM}$) with respect to the gas metallicity ($Z_{\rm gas}\equiv M_{\rm Z}/M_{\rm ISM}$). The PAH-to-dust mass ratios are also shown in Figure \ref{fig:dtog_Z} ({\it bottom}). 
In our standard case ($\tau_{\rm SF}=5$ Gyr and $\tau_\mathrm{DC}=3$ Myr), 
the PAH abundance slowly increases until the metallicity exceeds $\sim0.1~Z_\odot$, 
and the increase becomes rapid. 
At $Z_\mathrm{gas}\simeq0.3~Z_\odot$, the increase slows down again. 
This evolutionary change of the PAH abundance is associated with the evolution of the dust abundance. 
At low metallicities ($Z_\mathrm{gas}\la 0.1~Z_\odot)$, the dust abundance slowly increases by the production by SNe and AGB stars and starts increasing rapidly by grain growth at a certain metallicity called ``critical metallicity'' ($Z_\mathrm{cr}$) in \citet{asano12}. 
The critical metallicity is the switching point of the primary source of dust from stars (AGB stars and SNe) to the grain growth in the ISM, and for our standard case, $Z_\mathrm{cr}\simeq 0.1~Z_\odot$. 
The deceleration of the PAH abundance increase after $Z_\mathrm{gas}\simeq0.3~Z_\odot$ is due to the saturation of grain growth. 
That is, as most gas-phase metals have already been accreted onto dust, grain growth becomes ineffective.

Although the overall evolution of the PAH abundance is similar to that of the dust abundance, the PAH abundance increases more rapidly compared to the dust abundance. It is more clearly shown by the PAH-to-dust mass ratio (Figure \ref{fig:dtog_Z}, $bottom$): This ratio is very low at low metallicities and shows the drastic increase although the rapid increases of the PAH abundance and the dust abundance occur simultaneously. This is a natural consequence of PAH formation by shattering of carbonaceous grains: since shattering is a collisional process, it is sensitive to the dust abundance \citep{hiro09}. The details of each PAH formation and destruction process are looked into in the next subsection.

The lifetime of DC ($\tau_{\rm DC}=1$ or 10 Myr) has an appreciable influence 
on the evolution of the PAH abundance. Because $\tau_{\rm DC}$ affects 
the coagulation timescale ($\tau_{\rm coag}$) only (section \ref{sec:ctime}), 
the variation appears at the late stage of the PAH evolution when coagulation becomes efficient 
(see Figure \ref{fig:dtog_Z}). According to Equation (\ref{eq:tauc}), 
$\tau_{\rm coag}$ has a dependence on $\tau_{\rm DC}$. 
As the metallicity exceeds the critical metallicity, 
the dust-to-gas ratio increases rapidly because of dust growth, 
which makes $\tau_{\rm coag}$ decrease rapidly. 
When $\tau_{\rm coag,DC1}$ becomes shorter than $\tau_{\rm DC}$,
however, $\tau_{\rm coag}$ remains nearly steady because $\tau_\mathrm{DC}$ 
is adopted for the coagulation timescale in such a case 
as we explained in Section 2.1.2. For a long $\tau_{\rm DC}$ (10 Myr),
$\tau_{\rm coag}$ stops decreasing early and remains at a relatively high value. 
As a result, the effect of coagulation becomes limited (dotted lines in Figure \ref{fig:dtog_Z}). 
In contrast, a short $\tau_{\rm DC}$ (1 Myr) results in a short $\tau_{\rm coag}$ 
at the late evolutionary stage, so coagulation strongly affects the PAH abundance 
(solid lines in Figure \ref{fig:dtog_Z}).

\begin{figure}
\includegraphics[width=80mm]{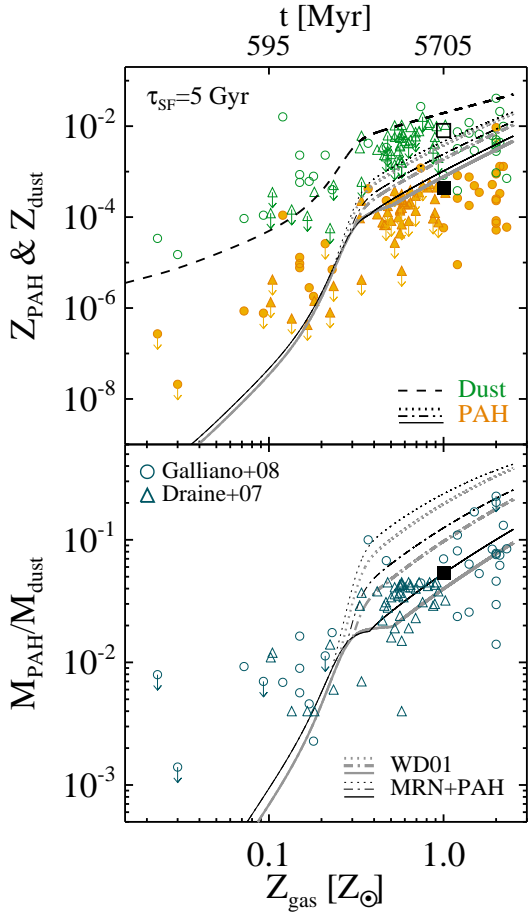}
\caption{Same as the case of $\tau_{\rm SF}=5$ Gyr in Figure \ref{fig:dtog_Z} 
overlaid with those calculated for the WD01 size distribution for comparison (thick grey lines).
\label{fig:dtog_wd01}}
\end{figure}

In addition, we investigate the effect of different star formation timescales 
($\tau_\mathrm{SF}=0.5$ or 50 Gyr).
For a shorter $\tau_{\rm SF}$, the evolution of the PAH abundance is more drastic.
The PAH abundance in the case of $\tau_\mathrm{SF}=0.5$ Gyr starts increasing at the higher metallicity ($Z_\mathrm{gas}\simeq0.4~Z_\odot$) than in the standard case but reaches the similar value finally. 
The difference results from the higher critical metallicity for the shorter $\tau_\mathrm{SF}$. 
The critical metallicity for $\tau_\mathrm{SF}=0.5$ Gyr is about 0.3 $Z_\odot$ \citep{asano12}, 
so the dust abundance as well as the PAH abundance increases rapidly when the metallicity in the galaxy becomes higher than the critical metallicity. In contrast, for a large $\tau_{\rm SF}$ (50 Gyr), 
the rapid increase of the PAH and dust abundances occur at a low metallicity ($Z_\mathrm{gas}\simeq0.04~Z_\odot$), 
which is consistent with the critical metallicity for $\tau_\mathrm{SF}=50$ Gyr ($Z_\mathrm{cr}\simeq0.03~Z_\odot$). 
Note that the metal enrichment in a galaxy with such a large $\tau_{\rm SF}$ takes much longer time than the other cases, 
and as the evolution of the dust abundance in \citet{asano12} terminates at 10 Gyr, 
the evolution for $\tau_{\rm SF}=50$ Gyr is stopped at the low metallicity.

In the case of $\tau_{\rm SF}=5$ Gyr, we compare the evolutions 
of the PAH abundance for the two size distributions, 
MRN+enhanced PAH (solid lines) and WD01 (dotted lines) in Figure \ref{fig:dtog_wd01}. 
The differences between the two cases are insignificant, 
and the case for WD01 has a slightly less PAH abundance than the case of MRN+enhanced PAH. 
This difference basically results from the different shattering and coagulation 
timescales (see section 2.1.1 and 2.1.2). 
In the case of WD01, the shattering timescale is longer than 
that for MRN+enhanced PAH because the shattering 
in WIM is somewhat less efficient (Figures \ref{fig:shat}--\ref{fig:shat_wd01} and Table \ref{tab:tau0}). 
(Although the shattering timescale in CNM becomes shorter, 
it has only minor effect.) 
At the same time, the coagulation timescale is slightly longer than that for MRN+enhanced PAH 
(Figures \ref{fig:coag}--\ref{fig:coag_wd01} and Table \ref{tab:tau0}). 
Combining the two effects, the case for WD01 produces a bit less PAH abundance during the evolution.

\subsection{Contribution from each process}

\begin{figure*}
\begin{minipage}{177mm}
\centering
\includegraphics[width=170mm]{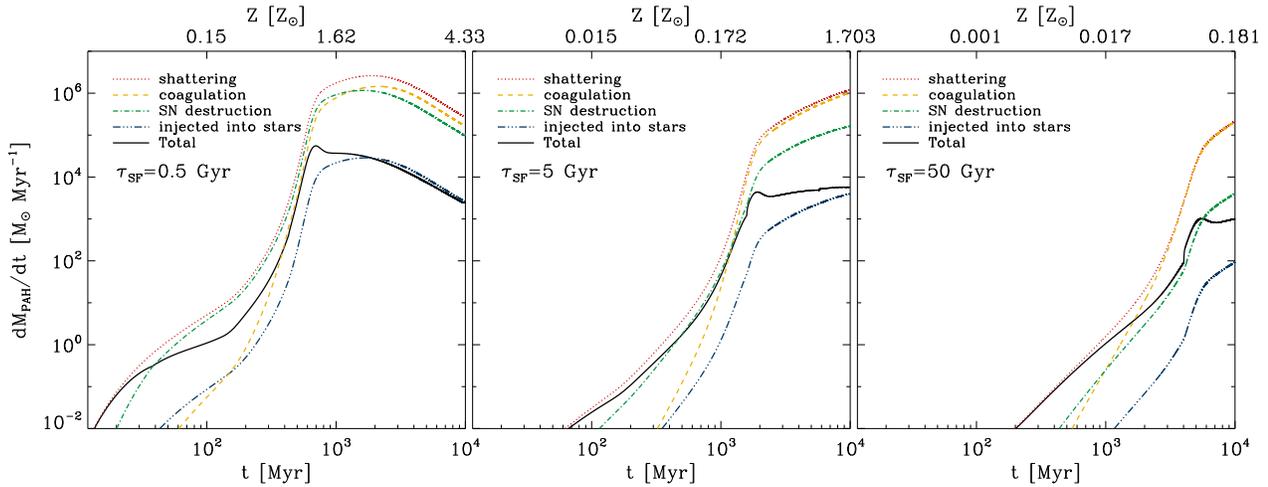}
\caption{
Evolution of each PAH formation/destruction rate along time or metallicity for $\tau_{\rm DC}=3$ Myr (equation \ref{eq:dmdt}). The star formation timescales ($\tau_{\rm SF}$) are set to be 0.5 Gyr (left), 5 Gyr (middle), and 50 Gyr (right). The individual rates of PAH formation by shattering, PAH coagulation, PAH destruction by SN shocks, and PAH injection into stars (astration) are represented with dotted, dashed, dash-dotted, and dash-double dotted lines, respectively. The total variation is shown with a solid line. 
\label{fig:dmdt_3Myr}} 
\end{minipage}
\end{figure*}

Now we examine how each of the PAH formation and destruction processes 
plays a role in the evolution of the total PAH mass. 
Figure \ref{fig:dmdt_3Myr} shows the time evolution of each PAH formation/destruction rate
corresponding to each term in the right hand side of Equation (\ref{eq:dmdt}) 
for $\tau_{\rm DC}=3$ Myr. In the case of $\tau_{\rm SF}=5$ Gyr (a standard case), 
the PAH production by shattering is inefficient in the early stage until $\sim500$ Myr ($Z_{\rm gas}\simeq 0.08~Z_\odot$). 
Afterward the shattering rate rapidly increases
due to the increase of the dust abundance by grain growth and slows down around 2 Gyr because grain growth is saturated by using up a large fraction of gas phase metals. 
The coagulation and SN destruction rates follow a similar trend 
because both processes are proportional to the PAH abundance. 
However, there is a weak trend that SN destruction is more dominant than coagulation in the early stage while the situation is reversed in the later stage.
This trend is explained by the fact that the coagulation rate also depends on the
abundance of carbonaceous dust, which nonlinearly increases with metallicity by
grain growth as explained in Section 3.1.
The effect of astration is minor over all evolutionary stages. 
Consequently, the total PAH mass changing rate is determined 
by the balance between shattering and SN destruction/coagulation. 
As the variations of the shattering and coagulation rates are comparable to each other after 2 Gyr, 
the total PAH mass changing rate becomes nearly constant. 

In the case of $\tau_{\rm SF}=0.5$ and 50 Gyr, the time evolutions of PAH formation/destruction rates show some differences from the standard case with $\tau_\mathrm{SF}=5$ Gyr (Figure \ref{fig:dmdt_3Myr}). For $\tau_{\rm SF}=0.5$ Gyr, the shattering rate evolves more quickly and slows down after $\sim700$ Myr. The SN destruction rate is larger than that of $\tau_{\rm SF}=5$ Gyr due to higher SN rate and is more dominant than coagulation until the very late stage. The coagulation rate also increases faster than the standard case. After all formation/destruction rates reach their peaks, they start to decrease, and $dM_{\rm PAH}/dt$ decreases eventually. In the case of $\tau_{\rm SF}=50$ Gyr, on the other hand, the evolutions of PAH processing rates slowly progress, which is mainly because dust enrichment in the ISM slowly occurs. The total PAH mass is also lower than the other cases. Coagulation becomes the primary destruction process in the relatively early stage as SN destruction is inefficient in this case.

\subsection{Comparison with observational data}

We compare our model calculations with the observational data taken from \citet{gal08} and \citet{draine07}. \citet{gal08} determined the PAH and dust abundances of 35 nearby galaxies using their models of UV-to-radio spectral energy distributions. Their sample includes various types of galaxies which cover a wide range of metallicities and star formation activities. Most of the sources were observed by ISOCAM (5 to 16.5 \micron) on board the $ISO$ satellite, and the rest of them were observed by $Spitzer$ IRS spectrometer (5.2 to 38 \micron). To constrain the stellar and the dust emission components, optical-to-NIR photometric fluxes ({\it U, B, V, R, I, J, H}, and $K$) and $Spitzer$ MIPS or $IRAS$ fluxes were also taken into account.
The oxygen abundance data is taken from Table 1 in \citet{gal08}, which is converted to the
metallicity by assuming the solar oxygen abundance to be $12+{\rm log(O/H)}_{\odot}=8.83$.
The other set of data is adopted from \citet{draine07} who derived dust masses and PAH abundances of 65 galaxies showing strong IR emission in the $Spitzer$ IRAC/MIPS bands among the SINGS galaxy sample \citep{kenn03}. In addition to the $Spitzer$ fluxes, they also considered available SCUBA $850$ \micron~photometric data for 17 galaxies. The metallicities of 65 galaxies were measured by using optical spectrophotometry \citep{mous10}.

In Figures \ref{fig:dtog_Z} and \ref{fig:dtog_wd01}, our model calculations ($\tau_{\rm SF}=5$ Gyr) show a good agreement with the observed dust and PAH abundances of galaxies with various metallicities. The low PAH abundances at low metallicities are consistent with the observed values although the PAH (and dust) abundances at high metallicities are slightly higher than the observational results. 
The slight overprediction of dust is inherent in \citet{asano12}'s models.
This overprediction comes from the dust and metal yield data adopted by them (Section \ref{sec:stime}). 
They also assume the sticking efficiency in grain growth to be unity (i.e., maximum). In addition,
\citet{asano12} do not consider the grain size distribution, which affects the efficiency of dust growth
\citep[e.g.,][]{hiro11}. 
However, any dust models available so far have more or less the same kind of
uncertainties, so we do not further fine tune our model calculation.

The most important feature in our models is the rapid rise of the PAH abundance
at a sub-solar metallicity. This is a natural consequence of the link between dust and
PAHs: PAHs form as a result of shattering of dust grains. The sharp rise is thus due to the
rapid increase of the dust abundance by dust growth. As \citet{asano12} note,
dust growth becomes dominant over the dust formation by stellar sources at the critical metallicity. 
Therefore, our most important prediction is that
the PAH abundance increases at a certain metallicity, which is almost similar to
the critical metallicity of dust growth (i.e., the metallicity above which the 
increase of dust abundance is dominated by dust growth).

\section{Discussion}

\subsection{General features of the models}

Our PAH evolution models naturally explain both the deficit of PAH abundance in low-metallicity galaxies and
the metallicity dependence of the PAH abundance in high-metallicity galaxies. This is mainly because the PAH formation/destruction mechanisms in our models have dependence on the dust abundance. Both shattering and coagulation do not efficiently occur at low metallicities because the dust abundance is low. 
At low metallicities, dust production is mainly dominated by stellar sources (SNe and AGB stars), 
whose dust yield does not strongly depend on metallicity \citep[Figure 1 in][]{asano12}. On the other hand,
dust growth by accretion in the ISM becomes efficient after the metallicity
exceeds the critical metallicity, dominating the dust production in the galaxy afterward.
This strong metallicity dependence of the dust production is reflected in the strong metallicity
dependence of the PAH abundance. The advantage of our models is that the strong
metallicity dependence of the PAH abundance is naturally explained.

The relatively steady increase of the PAH abundance at high metallicities can also be explained by the behavior of the dust mass growth. 
At sub-solar and solar metallicities, the dust growth becomes ineffective (or saturated) 
because of the depletion of gas-phase metals. 
At this stage, the shattering rate does not increase rapidly, so
the increasing rate of the PAH abundance becomes steady. Yet, the PAH abundance increases
as the metallicity increases since the dust-to-gas ratio continues to increase.

\subsection{Effects of other PAH formation and destruction mechanisms}\label{sec:dis_other}

Although our models can explain the relation between PAH abundance and metallicity, 
it cannot preclude possibilities that other PAH formation/destruction mechanisms affect this tendency. 
To interpret the paucity of PAH detection in low-metallicity galaxies, 
several explanations have been suggested \citep[e.g.,][]{gal08,ohall06}. 
\citet{gal08} showed that delayed injection of carbon dust produced by carbon-rich AGB stars into the ISM results in the low PAH abundance in low-metallicity galaxies. 
The time delay originates from the lifetime of AGB stars, 
which is around 100 Myr for the most massive AGB stars. 
This value ($\sim100$ Myr) corresponds to a metallicity of $\sim0.05~Z_\odot$, 
at which the PAH abundance starts rising in their models. 
However, as Figure \ref{fig:dmdt_3Myr} shows, the metallicity at a given age of a galaxy 
can vary considerably depending on its star formation timescale. 
Furthermore, the metallicity does not always reflect the age of a galaxy \citep[e.g.,][]{kunth00}. 
For example, BCDs are not chemically evolved (metal-poor), but most BCDs are certainly not young. 
Therefore, to attribute the evolutionary trend of PAH abundance with metallicity to the contribution from AGB stars itself, 
it may require an additional dependence that the PAH production by AGB stars is physically associated with metallicity.

On the other hand, the paucity of PAH detection in low metallicity galaxies may be attributed not to the intrinsic lack of PAH formation paths but to the relatively efficient PAH destruction. \citet{ohall06} find that the [Fe II] $\lambda 26$ \micron/[Ne II] $\lambda 12.8$ \micron~flux ratio decreases as the 7.7 \micron~PAH strength increases in 18 starburst galaxies. They also find a strong anti-correlation between the [Fe II]/[Ne II] ratio and the metallicity of the host galaxy. Since the [Fe II] line emission is related to supernova (SN) shocks, the strong anti-correlation between the [Fe II]/[Ne II] ratio and the PAH strength could indicate that the PAH destruction by the enhanced SN activity is the cause of the lack of PAH emission in the low-metallicity galaxies. The lack of PAH emission detected in Galactic SNRs supports the PAH destruction by SN shocks. However, recent observations reveal the existence of PAH emission in some SNRs \citep[e.g.,][]{seok12,tappe06}. In an SNR N49 in the Large Magellanic Cloud (LMC), 3.3 \micron~PAH features are detected where the SNR interacts with nearby molecular clouds \citep{seok12}. Another example, an SNR N132D in the LMC, shows a broad PAH plateau at 15 to 20 \micron~that is attributed to large PAHs \citep[$\ga4000$ carbon atoms,][]{tappe12,tappe06}. Again, the PAH plateau is detected where this SNR interacts with a molecular cloud. Therefore, PAH destruction in shocks is still uncertain especially when the shocks are interacting with molecular clouds.

The origin of the paucity of PAH emission in low-metallicity galaxies 
may be related to various effects of PAH formation/destruction 
mechanisms. Because the PAH processing in the ISM and the 
environmental effects on PAHs are not fully understood, 
the primary origin is not conclusive. However, as discussed in 
the previous section, our PAH evolution models with the primary 
PAH source being shattering of large carbonaceous grains 
give a natural explanation on both the deficit of the PAH 
emission in low-metallicity galaxies and the strong dependence 
of the PAH abundance on the metallicity in high-metallicity galaxies. 
To establish our scenario, observational evidence of PAH 
formation by shattering would be crucial. For example, 
\citet{miville02} presented the ISOCAM observations of the 
Ursa Major cirrus with {\sc Hi}, CO, and IRAS data. It is found 
that the MIR emissivity increases in a filament which shows a strong 
transverse velocity gradient whereas the MIR emissivity decreases in 
a CO emitting molecular gas. They considered that the variation of 
the MIR emissivity is mainly due to the change of the abundance of 
small dust grains and attributed it to shattering and coagulation in 
turbulence. We expect that PAH processing in turbulence can be 
further tested in such an environment by examining the variation of 
PAH features using both photometric and spectroscopic IR observations 
and by better constraining the grain size distribution from MIR-submillimeter 
observations with high spatial resolutions. 

In addition, theoretical studies on the PAH processing taking detailed chemical properties 
into account will provide better insight into the PAH evolution and its dependence on metallicity 
(see also the next subsection).


\subsection{Uncertainty in the models}\label{sec:unc}

Our models provide the first framework that enables us to treat the
evolution of PAH abundance over the entire galaxy evolution.
There are some possibilities of improving our framework if better
knowledge on PAH properties and more precise determination of
PAH abundance in galaxies are obtained in the future. We briefly
mention some possible future improvements.

We have adopted the material properties of graphite for PAHs.
However, the detailed properties of PAHs are not necessary as
far as our models are concerned, since the PAH formation rate is
determined by shattering of large carbonaceous dust, to which
we can safely apply the bulk properties. Moreover, coagulation
of PAHs is determined by the collision rate, which is insensitive
to the detailed material properties (the assumption
is that the sticking efficiency is unity). For the destruction by SN
shocks, we have taken into account the microscopic properties
of PAHs as explained in Section \ref{sec:sn}. Any further details of PAH
properties would not affect our results significantly.

\citet{jones09} suggested a-C:H as the major component of
carbonaceous dust grains. Because of its lower material
density, the efficiency of dust growth is higher for a-C:H
than for graphite by a factor of 2
\citep{hiro11}. At the same time, the destruction
timescale of a-C:H is shorter than that of graphite by a
factor of 3 \citep{serra08}, indicating
that the carbonaceous dust-to-gas ratio, which is
determined by the equilibrium between dust growth and
destruction \citep{inoue11} is insensitive to
the material adopted. The material
properties relevant to shattering are better known
for graphite \citep{jones96} than for a-C:H.
Indeed, the grain velocities that we adopted are calculated based on graphite
in \citet{yan04}. If more experimental data is accumulated
for a-C:H in the future, it would be interesting to calculate
the grain velocities based on the material properties of
a-C:H.

\section{CONCLUSION}

The paucity of PAH detection in low-metallicity ($Z\la 0.1~Z_{\sun}$) galaxies and the metallicity-dependence of PAH abundance in high-metallicity ($Z\ga 0.3~Z_{\sun}$) galaxies have been noted observationally for decades, but definite explanations for them have not been given. In this paper we developed a new framework for the evolution of the PAH abundance in the ISM on a galaxy-evolution timescale, assuming that PAHs form through fragmentation of carbonaceous grains \citep{jones96}. In our models, shattering in interstellar turbulence is considered as the primary PAH formation mechanism while coagulation onto large dust grains, destruction by supernova shocks, and astration are considered as the PAH destruction mechanisms. Shattering and coagulation are physically related to the dust abundance, 
the evolution of which is calculated based on the dust evolution model of \citet{asano12}.

To estimate relevant timescales for the evolution of the PAH abundance, we
numerically calculated shattering and coagulation in each ISM phase.
The shattering and coagulation timescales depend on the dust-to-gas ratio, and the coagulation timescale is also affected by the lifetime of a dense cloud. While shattering is the most effective in WIM with some contribution in CNM, coagulation is the most effective in DC. Taking the PAH formation and destruction mechanisms into account, we found that the calculated PAH abundance is low at low metallicities, increases rapidly beyond a certain metallicity, and reproduces the Milky Way value eventually. 
This trend shows a good agreement with observed PAH abundances of galaxies with a wide metallicity range.

The low PAH abundance at low metallicities and the rapid enhancement of PAHs at high metallicities in our models are a natural consequence of PAH formation and destruction mechanisms being associated with dust abundance. In metal-poor environments, PAH formation by shattering is ineffective due to the low dust abundance. Above a certain metallicity (critical metallicity), dust growth in the ISM becomes the main source of dust in galaxies, which is regulated by metallicity \citep{asano12}. As the ISM is enriched with
metals, PAH formation by shattering becomes effective because of the rapid increase of the dust abundance by the dust growth. Our models provide a coherent explanation to the observed trend of the PAH abundance in galaxies, supporting the validity of shattering and coagulation as the PAH formation and destruction mechanisms in the ISM.

\section*{Acknowledgements}
HH is supported through NSC grant 102-2119-M-001-006-MY3.
RSA acknowledges the support from the Grant-in-Aid for
JSPS Research under Grant No.\ 23-5514.

\clearpage

\label{lastpage}

\end{document}